# Solution-processed NiPS$_3$ thin films from Liquid Exfoliated Inks with Long-Lived Spin-Entangled Excitons

*Andrii Shcherbakov[1,2], Kevin Synnatschke[3,4], Stanislav Bodnar[1,2], Johnathan Zerhoch[1,2], Lissa Eyre[2,5], Felix Rauh[2], Markus W. Heindl[1,2], Shangpu Liu[1,2], Jan Konecny[6], Ian D. Sharp[2], Zdenek Sofer[6], Claudia Backes[4,7] and Felix Deschler[1,2,\*].*

*Affiliations:*

*1) Physical-Chemistry Institute, Heidelberg University, Im Neuenheimer Feld 229, 69120 Heidelberg*

*2) Walter Schottky Institute and Physics Department, Technical University of Munich, Am Coulombwall 4, 85748 Garching bei Munich, Germany*

*3) School of Physics, Trinity College Dublin, The University of Dublin, Dublin 2, Ireland*

*4) Applied Physical Chemistry, Heidelberg University, Im Neuenheimer Feld 253, 69120 Heidelberg, Germany*

*5) Electrical Engineering Division, University of Cambridge, 9 JJ Thomson Ave, Cambridge, CB3 0FA, United Kingdom*

*6) Department of Inorganic Chemistry, Faculty of Chemical Technology, University of Chemistry and Technology Prague, Technická 5, 166 28, Prague 6, Czech Republic*

*7) Physical Chemistry of Nanomaterials, University of Kassel, Heinrich-Plett-Straße 40, 34132 Kassel, Germany*

## Abstract

Antiferromagnets are promising materials for future opto-spintronic applications since they show spin dynamics in the THz range and no net magnetization. Recently, layered van der Waals (vdW) antiferromagnets have been reported, which combine low-dimensional excitonic properties with complex spin-structure. While various methods for the fabrication of vdW 2D crystals exist, formation of large area and continuous thin films is challenging because of either limited scalability, synthetic complexity, or low opto-spintronic quality of the final material. Here, we fabricate centimeter-scale thin films of the van der Waals 2D antiferromagnetic material NiPS$_3$, which we prepare using a crystal ink made from liquid phase exfoliation (LPE). We perform statistical atomic force microscopy (AFM) and scanning electron microscopy (SEM) to characterize and control the lateral size and number of layers through this ink-based fabrication.

Using ultrafast optical spectroscopy at cryogenic temperatures, we resolve the dynamics of photoexcited excitons. We find antiferromagnetic spin arrangement and spin-entangled Zhang-Rice multiplet excitons with lifetimes in the nanosecond range, as well as ultranarrow emission linewidths, despite the disordered nature of our films. Thus, our findings demonstrate scalable thin-film fabrication of high-quality $NiPS_3$, which is crucial for translating this 2D antiferromagnetic material into spintronic and nanoscale memory devices and further exploring its complex spin-light coupled states.

## Introduction

The discovery of graphene,[1] launched a new epoch in modern material science and provided a basis for entirely new classes of advanced optoelectronic devices. Since then, different families of 2D materials have emerged, including transition metal dichalcogenides (TMDs),[2] black phosphorus (BP),[3] and metal hexathiohypo diphosphates.[4] Each class of these van der Waals crystals[5] exhibits unique optoelectronic, magnetic, and mechanical properties in their low dimensional forms. More recently, the exploration of 2D magnetic materials has opened access to new applications in opto-spintronics due to enhanced spin fluctuations and magnetic anisotropy, which can also be tuned externally via doping, electric and magnetic fields, mechanical stress, or formation of Moiré patterns.[6–10] Among such materials, antiferromagnets are particularly promising candidates for spintronics because of their natural spin dynamics in the THz range.[11] These materials show zero net magnetization, which leads to the absence of stray magnetic fields, increasing the stability of magnetic order and leading to improved performance for memory applications.[12–15] One of the most promising classes of 2D antiferromagnets are the transition metal phosphorus trisulfides ($MPS_3$, where M can be Co, Fe, Mn, or Ni), which is attractive because the antiferromagnetic spin arrangement can be flexibly tuned by exchanging the metal atoms.[16–19] In addition, one can replace sulphur with selenium, which allows the electronic properties to be modified while preserving the crystal structure.[8]

$NiPS_3$ is a 2D antiferromagnetic layered metal phosphorus trisulfide, which is characterized by a hexagonal configuration of Ni atoms, with each Ni surrounded by six S atoms (Fig. 1a) to form trigonally distorted octahedra. Within one layer, three ligand atoms are located below and three above the $Ni^{2+}$ ion (Fig. 1b). In other words, the divalent $Ni^{2+}$ is surrounded by hexathiohypo diphosphate ions ($P_2S_6^{4-}$), forming a honeycomb lattice in monoclinic C2/m symmetry. As such, the material is often referred to as $Ni_2P_2S_6$ to describe the structure more accurately. However, we will choose $NiPS_3$ for simplicity. The bulk material is an antiferromagnet below a Néel temperature of 155 K. The magnetic moments of Ni are mainly in the plane of the layer. They are colinear with the *a* direction of the crystal and form ferromagnetic zig-zag chains along this crystal axis.[20,21] Bulk NiPS$_3$ was previously used for UV photodetectors,[22,23] field-effect transistors,[24] and humidity sensors with high selectivity.[25] Recently, micromechanically exfoliated NiPS$_3$ sheets have been reported to show non-trivial excitonic Zhang-Rice (ZR) singlet/triplet states,[26–30] in which a strong hybridization of Ni 3d and S 3p orbitals gives rise to localized charge transfer states.[26,31] The ZR-states are created by the coherent many-body behavior of the spin and electron subsystems of the crystal, which creates an unusual link between magnetic and optoelectronic properties.[26,27,32–37] The luminescence from the ZR-excitons in NiPS$_3$ shows an extremely sharp linewidth for spontaneous emission (FWHM = 0.4 meV).[27] However, the exact nature of the ZR-excitons and the origin of the luminescence remain to be fully understood, and further experimental insights on the transient behavior of the spin and charge carriers of these spin-entangled excitons are desired.

Despite the intense scientific interest and the realization of novel physics with 2D magnetic compounds, the scalable fabrication of high quality low-dimensional materials remains challenging and is crucial for future device applications.[38] We now demonstrate that large area thin films of NiPS$_3$ nanosheets can be fabricated using 'inks' from liquid-phase exfoliation (LPE),[39] while also retaining stable Zhang-Rice excitons with extremely narrow linewidth comparable to previously reported micromechanically-exfoliated flakes.[26–30] We use the prepared thin film samples in combination with low temperature optical spectroscopy to gain insights into the energetics, interactions, and recombination dynamics of spin-entangled ZR-excitons and find unexpectedly long lifetimes on the nanosecond scale. Our demonstration has significant potential for enabling future technologies, since it provides high quality functional material and is compatible with a wide range of related antiferromagnetic VdW crystals.

## Results

**Ink preparation and thin film fabrication**

The schematic illustration of the ink preparation for the investigated samples of LPE NiPS$_3$ flakes is shown in Fig. 2a. The stock dispersion containing NiPS$_3$ nanosheets for fabrication of thin films was prepared by bath sonication (Branson, CPX2800-E, 130W) of ground NiPS$_3$ crystals in an inert atmosphere for 7 h. For this purpose, 250 mg of the ground material was immersed in 100 ml of dry and degassed isopropanol in a 250 ml round bottom flask. The stock dispersions were size-selected by liquid cascade centrifugation

(LCC) in multiple steps of iteratively increasing centrifugation speeds. Since 2D NiPS$_3$ has potential magnetic applications because of its distinctive spatial anisotropy of spin and charge subsystems, it is crucial to control the alignment of the nanosheets upon deposition onto the substrate. To this end, the NiPS$_3$ nanosheets were deposited in parallel alignment with respect to the substrate surface using a Langmuir Schaefer-type approach[40–42] in a custom-built setup (see Methods), which provides a high degree of in-plane alignment on glass and silicon substrates. As shown in Fig. 2b and 2c, respectively, the SEM images of prepared samples with large flakes (size selected via two-step 0.6-3k *g* centrifugation, for details on the sample labeling see Methods) and small flakes (10-30k *g)* demonstrate the homogeneous coating of substrates with tiled networks of NiPS$_3$ nanosheets covering several square centimeters. These two types of samples with large and small flakes will further be investigated in optical measurements, as discussed below.

To evaluate the detailed distribution of the number of layers and lateral sizes of the nano-flakes in the two samples, statistical atomic force microscopy (AFM) was used. To this end, diluted nanosheet inks were flash evaporated on silicon wafers (for further details on the process, see Methods).[39] The average thickness distribution for the sample containing small nanosheets is 5.5±0.24 layers (Fig. 2d), with individual flakes possessing lateral dimensions of 112±4 nm (Fig. 2e). Samples with large flakes show a broader size and thickness distribution, with an average thickness of 7.8±0.5 layers (Fig. 2d) and 255±10 nm lateral size (Fig. 2e). We note that for LPE, the number of layers is connected to the lateral size of the nanomaterial (Supporting Information Fig. 2).[43]

**Low-Temperature Photoluminescence of NiPS$_3$ Nanosheet Films**

To demonstrate that the inherent optoelectronic properties of our NiPS$_3$ nanosheet films are preserved after exfoliation, we investigated their luminescence with steady-state photoluminescence (PL). Importantly, we observe extremely narrow photoluminescence peaks below the Néel temperature in LPE NiPS$_3$, in line with previous reports on micromechanically exfoliated NiPS$_3$ crystals[27–30] and demonstrating that complex many-body magnetic state-dependent interactions can be achieved. In particular, high-resolution PL spectroscopy on the deposited LPE NiPS$_3$ nanosheets at 4K, excited with a 532 nm diode laser at a power density of 0.32 W/cm$^2$, shows an emission feature located at 1.4762 eV (Fig. 3a, 'Peak I'). The emission profile of this peak is asymmetric and non-Gaussian. Therefore, it was fitted with three Gaussian profiles at 1.4762 eV (peak A), 1.4775 eV (peak B), and 1.4790 eV (peak C), in agreement with literature.[27] All three subpeaks have a small full width half maximum (FWHM) of approximately 1.7 meV, suggesting that this emission stems from coherent excitations or localized defect states, in line with previously reported data.[27] We note that the FWHM of LPE NiPS$_3$ is around 5 times larger than reported for bulk NiPS$_3$ and coincides with the values for couple layer sheets which were micromechanically exfoliated.[27,28,30] We associate this with additional inhomogeneities coming from the edges of flakes[44–47] together with the non-zero size and thickness distribution, which can influence the bandgap energy[48] and energy of the exciton,[27] thereby causing emission broadening. Nevertheless, the FWHM is extraordinarily small considering the facile solution-based preparation of disordered 2D material nanosheet thin films. We further find PL in the range of 1.42-1.47 eV (Fig. 3b), which is likely associated with the phonon sideband of excitonic peak A. [28]

Samples containing large and small flakes of LPE NiPS$_3$ were characterized by similar PL lineshapes. However, some small changes, such as increased relative intensity of the subpeaks B and C and the emergence of a weak feature to the left of subpeak A for the small flakes, suggest a difference in the quantum confinement and influence from the edges, both of which should be more significant for the sheets with the smaller lateral size. Previous studies have shown that such narrow PL is observed down to 3-layered samples, in which the antiferromagnetic order is still preserved, with a slight blue shift for thinner sheets.[16,27,28] Thin films from large nanosheets showed an eight times enhanced PL intensity compared to films made from smaller sheets, which is in qualitative agreement with previous studies of PL dependence on the crystal thickness,[27,30] and attributed to increased number of emitters due to thicker films.

The luminescence from Peak I vanishes with increasing temperature (Fig. 4c and Supporting Information Fig. 5a), which is in line with previous reports[27] and can be considered as a result of the antiferromagnetic-to-paramagnetic phase change in the NiPS$_3$ thin films. However, the emission already disappears at approximately 100 K, which is below the Néel temperature of NiPS$_3$. Thus, we identify an additional effect from thermal heating on the excitonic luminescence in LPE NiPS$_3$, beyond the reported connection to the presence of antiferromagnetic order. The blue shift of the subpeaks (A, B, and C see Supporting Information Fig. 5a) upon cooling is typical of the temperature-dependent bandgap energy[49] change in semiconductors. The energy difference between subpeaks remains almost unchanged while heating the sample, though slight variations could be a consequence of minor softening of the responsible phonon bands[4] for material at elevated temperatures. Subpeak A is broadened at higher temperatures (Supporting Information Fig. 5a.) because of the increased phonon density, which results in stronger electron-phonon scattering. Importantly, subpeaks B and C also broaden with increasing temperature and follow a similar trend. One may hence conclude that they originate from the same transition as subpeak A. Furthermore, the temperature-dependent PL spectra show an additional narrow peak located at 1.4968 eV (Supporting Information Fig. 4, 'Peak II'). The peak is barely visible at 4 K, reaches its intensity maximum around 55 K, and disappears above 100 K. Previous studies have assigned this peak with phonons coupled to the exciton.[4,28]

These observations suggest three main processes acting on the emission from Peak I and Peak II: *a.* The existence of the antiferromagnetic ordering of the material; *b.* Thermal activation of the state driving emission of Peak II between 50 K and 60 K, which might be evidence of its indirect nature; and *c.* Destructive phonon scattering at higher temperatures which is a competitive process to *b.* According to the Zhang-Rice nature[26,33–35,37] of the ground and excited states underlying the investigated optical transitions, they usually involve a spin flip, which should lead to a certain degree of polarization of the emitted PL with respect to the crystal axis direction of the nanosheets.[27,28,30] We also measured polarization-dependent PL and did not find any preferable polarization direction, which can be attributed to the random orientation of the tiled $NiPS_3$ sheet network on the substrate plane. Nevertheless, these luminescence measurements on our solution-fabricated nanosheet thin films allow us to conclude that antiferromagnetic order is achieved at cryogenic temperatures and that Zhang-Rice excitons are formed with optoelectronic properties comparable to the ones reported for micromechanically-exfoliated sheets.

**Low-temperature broadband transient absorption of $NiPS_3$**

Having characterized the excitonic energetics and interactions via steady state PL measurements, we now move to the transient exciton dynamics in our $NiPS_3$ nanosheet thin films, which were studied using with low-temperature transient absorption (TA) spectroscopy. For this purpose, we excite above the bandgap with a 2.41 eV pump laser beam (275 fs pulse duration, 5 kHz repetition rate). Broadband probe beams were produced in $CaF_2$ to detect transmission properties through the samples in the region of the excitonic transitions.

The probe energy vs. time delay TA map of excitonic transitions measured at 4 K (Supporting Information Fig. 6) shows distinct positive, long-lived TA signals (bleach signals) at 1.4778 eV and 1.4974 eV, which align well with the positions of Peak I and Peak II in our PL data. We assign the strong negative signal at early times to a cross-phase modulation[50] artifacts when both pump and probe pulses overlap in the sample causing nonlinear effects. We note that the detected TA excitonic peaks only exist in the presence of both the pump and probe beams, which excludes a contribution of the PL signal. We further find a broad positive signal over the whole region of probe energies, which we associate with the low energy shoulder of the previously reported 1.7 eV d-d transition dressed with phonon replicas. [51]

The temperature-dependent TA of both excitonic peaks shown in Fig. 4 and Supporting Information Fig. 8 vanish above 120 K, indicating their dependence on antiferromagnetic order. The integrated intensity of Peak II drops faster than Peak I while heating the sample. Together with the observation of more substantial broadening, these findings suggest that the Peak II exciton-phonon bound state scatters more effectively with other phonons than the Peak I excitonic state.

We analyze the kinetics of the excitonic Peaks I and II up to 20 ps after the pump excitation (Fig. 4b). The peaks are overlapped with the cross-phase modulation[50] background at early times and the phonon replica background during all probed time delays.[51] We obtain the pure kinetics of the excitonic Peaks I and II (black lines Fig. 4b) by subtracting the kinetics of the background, taken from the probe region of 1.46±0.01 eV for Peak I and 1.51±0.01 eV for peak II (blue lines Fig. 4b), from the as-measured convoluted excitonic-background kinetics (green lines Fig.4 6b). From single-exponential decay fits we determine fast kinetics of the backgrounds $\tau_1$ (for Peak I) and $\tau_3$ (for Peak II) that lay in the range 3.5 – 4.8 ps, which is comparable to the decay of the ground state bleach (GSB) region (Supporting Information Fig. 6 & 7).

After the background correction we can only identify a short lifetime component of Peak II, which is equal to $\tau_2$= 99 ps. The remaining decay kinetics of excitonic Peak I and Peak II are beyond the time range of our setup, in the nanosecond range. The long lifetime of these states is in contrast to the reported PL lifetime[28] (on the order of 10s of ps) and the kinetics of transient reflection[51] (below 1 ps). We attribute this difference to the fact that our nanosheet films now provide access to optical processes that unambiguously probe the electronic absorption of photoexcited populations, and we resolve that the decay times of the photoexcited excitonic populations must be unexpectedly long, into the nanosecond range.

## Conclusion

Our report of spin-entangled excitonic Zhang-Rice states with narrow luminescence linewidth, as well as long-lived transient absorption, in thin films fabricated from LPE NiPS$_3$ nanosheet inks, demonstrates a scalable, solution-based fabrication route for thin films of magnetic 2D materials, which preserves a high purity of the antiferromagnetic order. The evidence of excited states responsible for Peak I and Peak II transitions have previously been reported in a resonant inelastic X-ray spectroscopy (RIXS) study and was corroborated by a theoretical model.[27] For the luminescence from Peak II, different origins have been proposed, including two-magnon-exciton coupled states,[27] exciton-phonon coupled states[28] or formation of many-body Zhang-Rice quasiparticles, as reported for cuprates.[26,32–37] From our temperature-dependent PL analysis of this peak, and the energy spacing to Peak I, we assign the origin of Peak II to coupling to the phonon mode at 180 cm$^{-1}$, which is highly spin-dependent and weakly pronounced in monolayer nanosheets.[4,18]

The transition between local Zhang-Rice singlet and triplet states is further highly sensitive to antiferromagnetic order and scattering with phonons in the system, which is a competing process in forming the Peak I transition. A comparison of our PL and TA results suggests that the Peak II transition has a higher oscillator strength than Peak I, but that PL emission occurs mainly from the lower-energy Peak I transition. From the TA kinetics, we conclude that the lifetime of the Peak II transition is shorter than the Peak I transition. Together with stronger scattering observed in temperature-dependent TA, we confirm that the Peak II excited state is based on the Peak I excited mode.

In conclusion, LPE inks of 2D magnetic semiconductor nanosheets present a facile, scalable thin film fabrication process that retains the excitonic properties of the studied 2D magnetic materials. Access to thin film samples of this promising class of materials will be instrumental in investigation of the detailed physical origins of Peak I and Peak II, and the translation of these materials into opto-spintronic applications.

## Methods

**Solvent preparation.**

Isopropanol (IPA) was dried over molecular sieves (10 Å, dried at 300 °C in vacuum overnight) for 2 days, distilled and saturated with nitrogen prior to use.

**Liquid Phase Exfoliation (LPE).**

The stock dispersion containing NiPS$_3$ nanosheets was prepared by bath sonication (Branson, CPX2800-E, 130W) of ground NiPS$_3$ crystals in an inert atmosphere for 7 h. For this purpose, 250 mg of the ground material was immersed in 100 ml of dry and degassed isopropanol in a 250 ml round bottom flask. In order to avoid effects from sample heating, the water in the bath was kept below 10 °C by exchanging it with ice-cooled water every 30 min. The atmosphere inside the flask was kept inert by constantly bubbling nitrogen through the solvent. Dispersions prepared according to this protocol are referred to as stock dispersion.

**Size selection.**

The stock dispersions were size selected by liquid cascade centrifugation (LCC) in multiple steps of iteratively increasing centrifugation speeds. A Hettich Mikro220R centrifuge equipped with a fixed-angle rotor (1195A) was used for this purpose. Each centrifugation step was performed for 2 h at 20 °C. Four different nanosheet sizes were isolated after sedimentation of the material at 0.6, 3, 10, and 30k $g$. After each step, the sediment was redispersed for analysis in a reduced solvent volume (~1 mL IPA), and the supernatant was used for the subsequent step. The expression "0.6–3k $g$" denotes consecutive centrifugation steps. In this example, the sample was obtained by sedimentation of nanosheets at 3000 $g$ from the supernatant obtained after centrifugation at 600 $g$.

**Thin Film Formation.**

Nanosheets were processed into thin films by a modified Langmuir Schaefer approach, using a custom-built setup. To this end, a substrate was immersed in deionized water and horizontally aligned to the liquid level. Approximately 2 ml of hexane was added to the water surface and concentrated nanosheet dispersion was added to the interface between water and hexane until a closed film was observed. The substrate was then vertically moved through the nanosheet film and left to dry.

**Atomic Force Microscopy (AFM).**

For AFM measurements, a Dimension ICON3 scanning probe microscope (Bruker AXS S.A.S.) was used in ScanAsyst mode (non-contact) in air under ambient conditions using aluminum-coated silicon cantilevers (OLTESPA-R3). The concentrated dispersions were diluted with isopropanol to optical densities <0.1 at a wavelength of 300 nm. A drop of the dilute dispersions (15 µL) was flash-evaporated on pre-heated (175 °C) Si/SiO$_2$ wafers (0.5x0.5 cm$^2$) with an oxide layer of 300 nm thickness. After deposition, the wafers were rinsed with ~15 mL of water and ~15 mL of isopropanol and dried with compressed nitrogen. Typical image sizes ranged from 20x20 µm$^2$ for larger nanosheets to 5x5 µm$^2$ for small nanosheets and were acquired at scan rates of 0.5-0.8 Hz with 1024 lines per image. Previously published length corrections were used to correct lateral dimensions from cantilever broadening.

**Scanning Electron Microscopy (SEM).**

SEM images were acquired with a JEOL JSM-7610F field emission scanning electron microscope (FE-SEM), using an in-lens Schottky field emission electron gun with different acceleration voltages but not higher than 15 kV at 2.5×10$^{-9}$ mbar. The images were measured with a dual (upper and lower) detector system consisting of collector-, scintillator-, light guide-, and photomultiplier units for secondary electron imaging (see Supporting Information Fig. 1).

**Steady-state Photoluminescence Measurements**

The steady-state PL was measured on the LPE NiPS$_3$ placed on the Si substrate in an open cycle He cryostat in reflection geometry. Excitation was performed with a 532 nm diode laser and detection was achieved with a high-resolution CCD spectrometer (HORIBA).

**Transient Absorption Measurements**

Transient absorption (TA) measurements were performed by exciting our LPE NiPS$_3$ samples on a transparent glass substrate with a 275 fs duration linearly polarized 514 nm pump laser beam (2$^{nd}$ harmonic of a diode-pumped Yb femtosecond laser Hiro + Pharos, Light Conversion). Probing of transient absorption in the visible range was done with the fundamental beam (1030 nm) of the laser-focused into a YAG crystal, which generated a broad spectrum of visible light. Probing of transient properties of the investigated samples in the NIR region was accomplished with the same laser-based noncolinear optical parametric amplifier (NOPA) and 2$^{nd}$ harmonic generation (Pharos + Orpheus + Prism compressor + Lyra, Light Conversion) of a 840 nm single color beam of 80 fs pulse duration, which was used for NIR white light generation in a calcium fluoride crystal. The crystal was vibrated on a loudspeaker to avoid its damage. Both pump and probe beams were directed through the front window of a closed cycle cryostat (Montana Instruments), spatially overlapped on the sample, and exit through a second window, after which the pump beam was blocked. The probe beam was detected by a fast CMOS camera triggered by the laser. By varying the optical path of the pump pulse with a mechanical delay stage, we controlled the time between sample excitation and probing. A chopper wheel was used to modulate the pump beam by blocking every second pulse, thereby allowing the camera to detect probe pulses when the pump pulse was both on and off. Finally, the difference between pumped and non-pumped signals was found and divided by the non-perturbated transmission spectrum.

## Data availability

Data which was used in the evaluation of the conclusions of our findings is available from the Wiley Online Library or from the authors upon reasonable request.

## Acknowledgments

This project has received funding from the European Research Council (ERC Starting Grant agreement no. 852084 — TWIST). F. D acknowledges financial support from the Deutsche Forschungsgemeinschaft (DFG) under the Emmy Noether Program (Project 387651688). I.S. acknowledges DFG under Germany´s Excellence Strategy – EXC 2089/1 – 390776260.

## Supporting Information

Supporting Information which includes detailed sample morphology analysis, temperature dependencies with fitting parameters for them and TA maps is available.

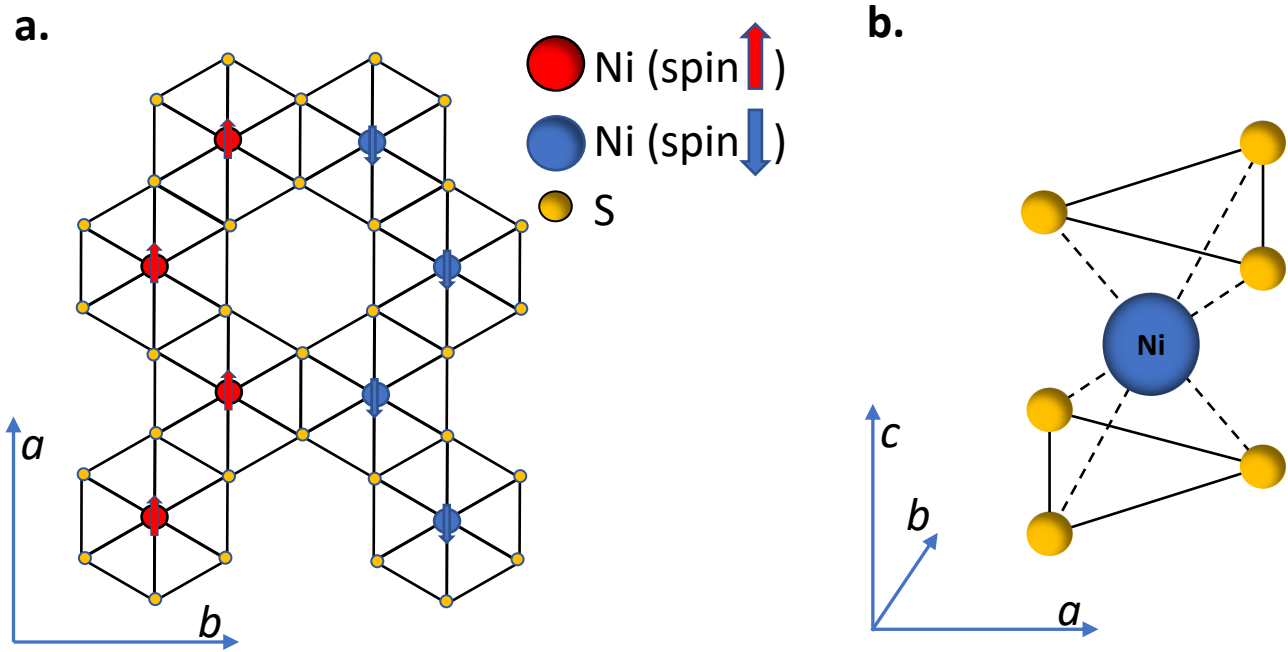

**Fig. 1. Crystal structure of NiPS$_3$ antiferromagnetic phase.** Red (blue) circles represent Ni atoms with spins parallel (antiparallel) with crystal axis *a*, and yellow circles represent S atoms. (a) Top view of the NiPS$_3$ sheets, which are parallel to the *ab* plane of the crystal lattice. (b) View of the local environment of the Ni ion surrounded by 6 S ligands.

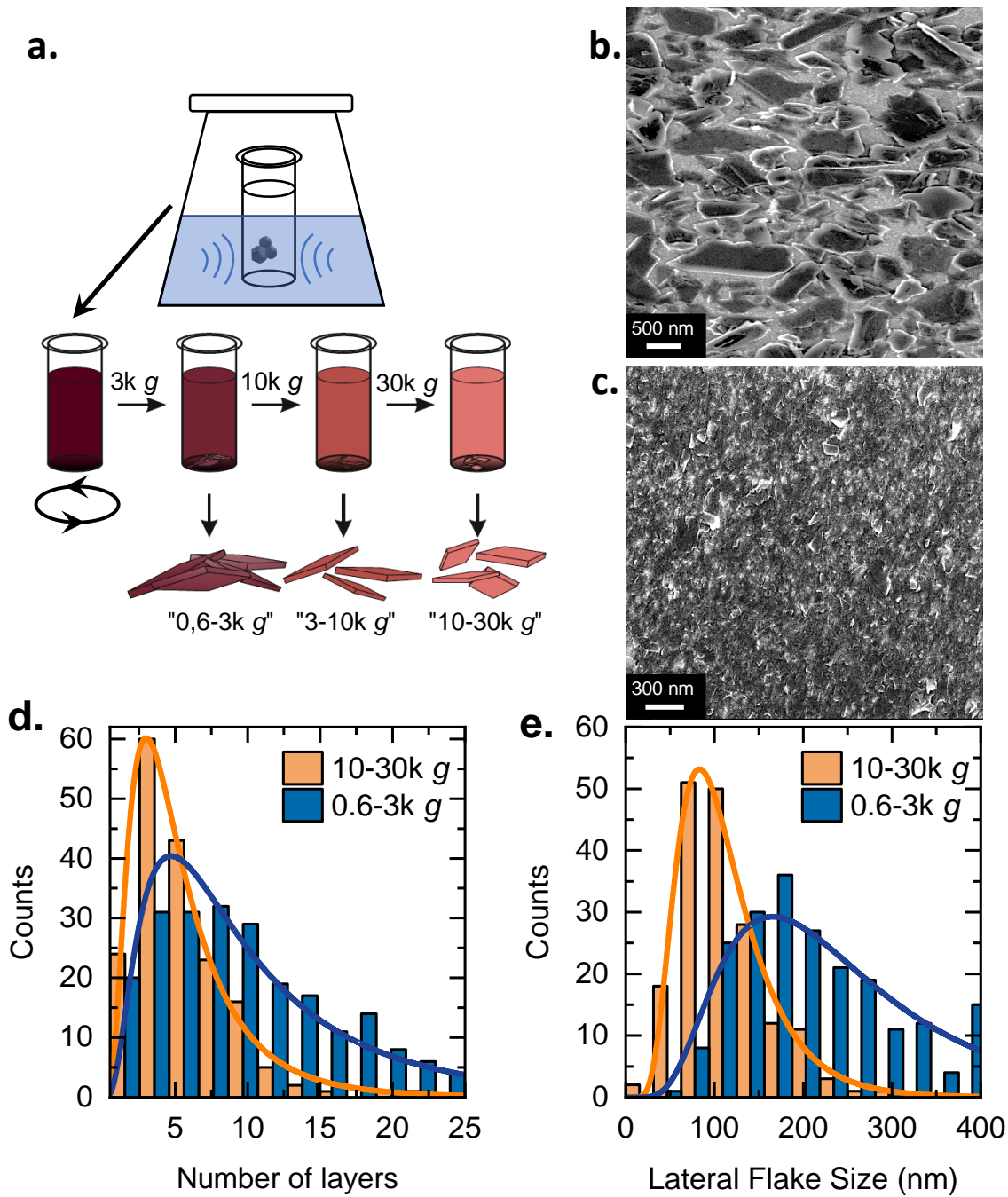

**Fig. 2. Liquid-phase exfoliation of NiPS$_3$ and statistics of LPE NiPS$_3$ nanosheet sizes.** (a) Schematic illustration of the ink fabrication for thin film preparation. The bulk material is ultrasonicated in the ultrasonic bath, followed by step-wise centrifugation for flake size selection. (b,c) SEM images of thin films from Langmuir-Blodgett deposition for large (b) and small (c) nanosheets of LPE NiPS$_3$. (d) Distribution of layer number for small (10-30k $g$) and large (0.6-3k $g$) nanosheets of LPE NiPS$_3$. (e) Distribution of lateral sizes of small and large nanosheets. The lines are guides to the eye.

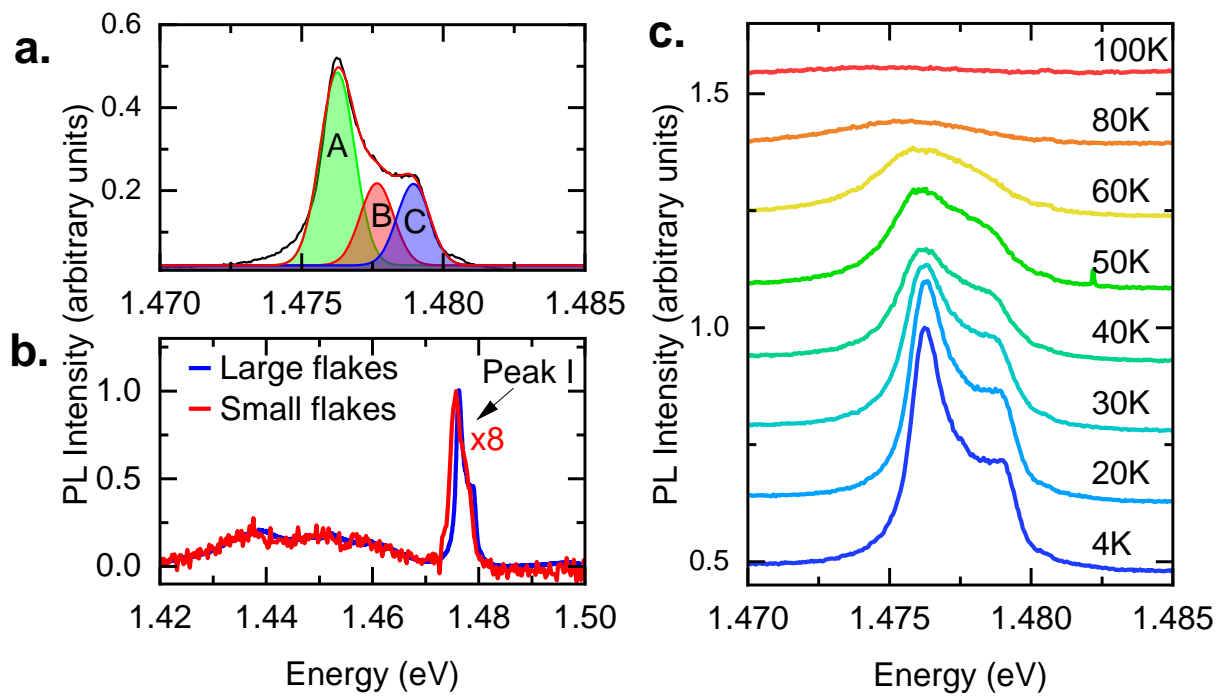

**Fig. 3. Photoluminescence fine structure of LPE NiPS$_3$.** (a) Fine structure of excitonic PL spectra of large nanosheets (0.6-3k *g*) at 4K under 532 nm laser excitation (black line) with fit (red line) to three Gaussian subpeaks A (green), B (red), and C (blue) of Peak I. (b) Excitonic PL showing similar spectra independent of nanosheet size. (c) Temperature-dependence of excitonic PL of large nanosheets (0.6-3k *g*) showing an increase of PL intensity below the Néel temperature.

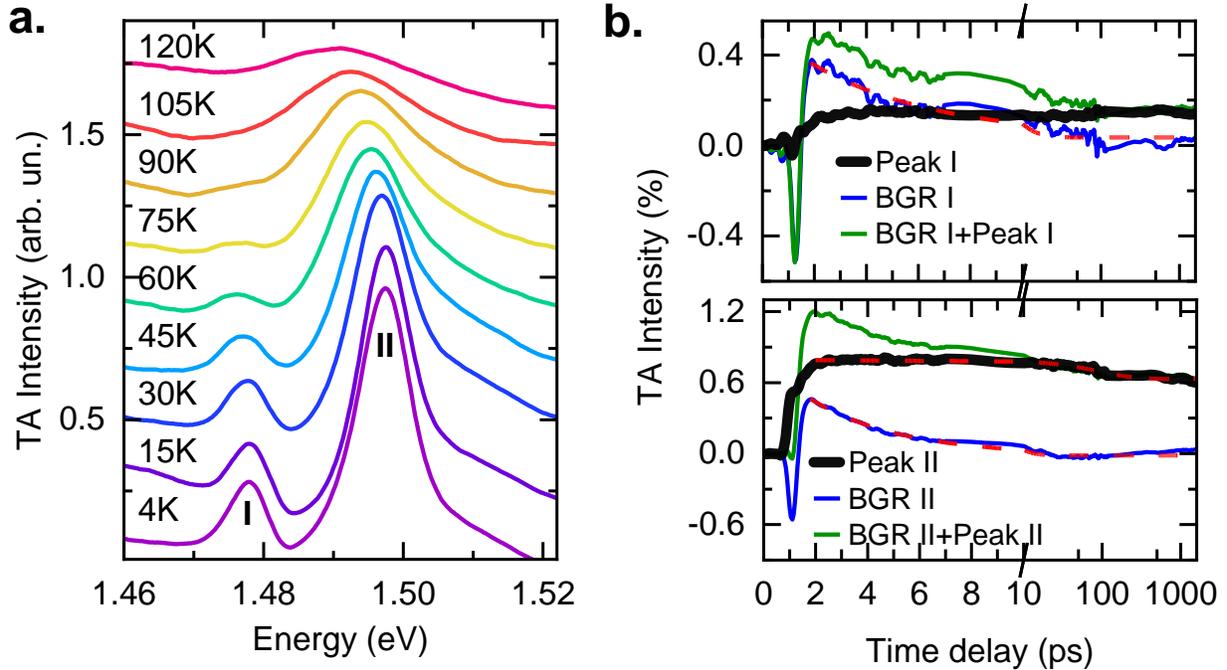

**Fig. 4. Transient absorption excitonic transition dynamics of large (0.6-3k *g*) nanosheets of LPE NiPS$_3$.** (a) Temperature-dependent ΔT/T spectra measured from 4 K to 120 K in the region of excitonic absorption of LPE NiPS$_3$, averaged over the time delay between 20 – 150 ps. We find excitonic peaks at 1.478 eV (Peak I) and 1.497 eV (Peak II). (b) Kinetics of excitonic Peak I (upper graph) and Peak II (lower graph), obtained by subtracting the kinetics from a broad superimposed background signal (BGR I and II), as shown. Single exponential fits to the experimental data (red dashed lines) with time constants $\tau_1$=4.8 ps (BGR I), $\tau_2$=99 ps (Peak II), and $\tau_3$=3.5 ps (BGR II).

# Supporting information

# Solution-processed NiPS$_3$ thin films from Liquid Exfoliated Inks with Long-Lived Spin-Entangled Excitons

*Andrii Shcherbakov[1,2], Kevin Synnatschke[3,4], Stanislav Bodnar[1,2], Johnathan Zerhoch[1,2], Lissa Eyre[2,5], Felix Rauh[2], Markus W. Heindl[1,2], Shangpu Liu[1,2], Jan Konecny[6], Ian D. Sharp[2], Zdenek Sofer[6], Claudia Backes[4,7] and Felix Deschler[1,2,*].*

*Affiliations:*

*1) Physical-Chemistry Institute, Heidelberg University, Im Neuenheimer Feld 229, 69120 Heidelberg*

*2) Walter Schottky Institute and Physics Department, Technical University of Munich, Am Coulombwall 4, 85748 Garching bei Munich, Germany*

*3) School of Physics, Trinity College Dublin, The University of Dublin, Dublin 2, Ireland*

*4) Applied Physical Chemistry, Heidelberg University, Im Neuenheimer Feld 253, 69120 Heidelberg, Germany*

*5) Electrical Engineering Division, University of Cambridge, 9 JJ Thomson Ave, Cambridge, CB3 0FA, United Kingdom*

*6) Department of Inorganic Chemistry, Faculty of Chemical Technology, University of Chemistry and Technology Prague, Technická 5, 166 28, Prague 6, Czech Republic*

*7) Physical Chemistry of Nanomaterials, University of Kassel, Heinrich-Plett-Straße 40, 34132 Kassel, Germany*

**Fitting of temperature-dependent integrated intensity of PL and TA excitonic peaks.**

The Arrhenius relation was used to analyze the temperature-dependent integrated intensity of the investigated excitonic peaks. In the classical form, such a relation is given by:

$$I = \frac{I_0}{1+\sum_{i=1}^{n}\left(A_i \exp\left(-\frac{E_i}{k_B T}\right)\right)}, \quad (1)$$

where $I_0$ is an integrated intensity of the peak at 0 K, $n$ is a number of recombination channels, $A_i$ is coupling constant, $E_i$ is the characteristic activation energy, $k_B$ is the Boltzmann constant, and $T$ is the temperature[1–3]. To simplify the fitting process, we put the formula above in a form:

$$y = \sum_{i=1}^{n}\left(A_i \exp(-E_i x)\right), \quad (2)$$

where $x = \frac{1}{k_B T}$ and $y = \frac{I_0}{I} - 1$, thereby allowing us to fit the data with a number, *n*, of exponential decays.

**Fitting of temperature dependent integrated intensity of PL and TA.**

In order to analyze PL and TA peak shifts as a function of the temperature, we use Bose-Einstein relation in the form:

$$E_i = E_{i\beta}(0) - \alpha_{i\beta}\left(1 + \frac{2}{exp\left(\frac{\theta_{i\beta}}{T}\right)-1}\right), \qquad (3)$$

Where $E_i$ is a temeperature dependent peak energy, $E_{i\beta}(0)$ is peak energy at absolute zero, $\alpha_{i\beta}$ represents an electron-phonon interaction and $\theta_{i\beta}$ is an average phonon temperature[2,4].

**Fitting of temperature dependent FWHM of PL and TA peaks.**

The full width at half maximum can be fitted with Bose-Einstein relation in the form:

$$\Gamma_i = \Gamma_{i0} + \frac{\Gamma_{iLO}}{\left(exp\left(\frac{\theta_{iLO}}{T}\right)-1\right)}, \qquad (4)$$

where $\Gamma_{i0}$ is the line shape broadening due to electron-electron interactions and Auger processes, $\Gamma_{iL0}$ is broadening from interaction between electrons and longitudinal optical phonons, and $\theta_{iLO}$ represents the longitudinal optical phonon temperature[2].

| Peak | $A_{1,2}$ | $E_{1,2}$ (meV) | $E_{i\beta}$ (eV) | $\alpha_{i\beta}$ (meV) | $\theta_{i\beta}$ (K) | $\Gamma_{i0}$ (meV) | $\Gamma_{iL0}$ (meV) | $\theta_{iLO}$ (K) |
|---|---|---|---|---|---|---|---|---|
| PL Peak I Subpeak a | 147406±77531; 0.7172±0.5396 | 4±3; 81±5 | 1.484±0.002 | 8±2 | 249±18 | 1.47±0.12 | 11.9±3.4 | 105±16 |
| PL Peak I Subpeak c | 13662±25737 | 0.2±0.6; 44±10 | - | - | - | - | - | - |
| TA Peak I | 31±20 | 26±6 | 1.479 | 1.2±0.2 | 56±10 | - | - | - |
| TAPeak II | 5.88±3.22; 53912±26085 | 8.1±1.9; 54.3±3.3 | 1.501 | 3.3±0.4 | 94±12 | 8.58±0.33 | 19±8 | 115±31 |

Table 1. Fitting Parameters for temperature-dependent PL and TA excitonic peaks

Large Nanosheets

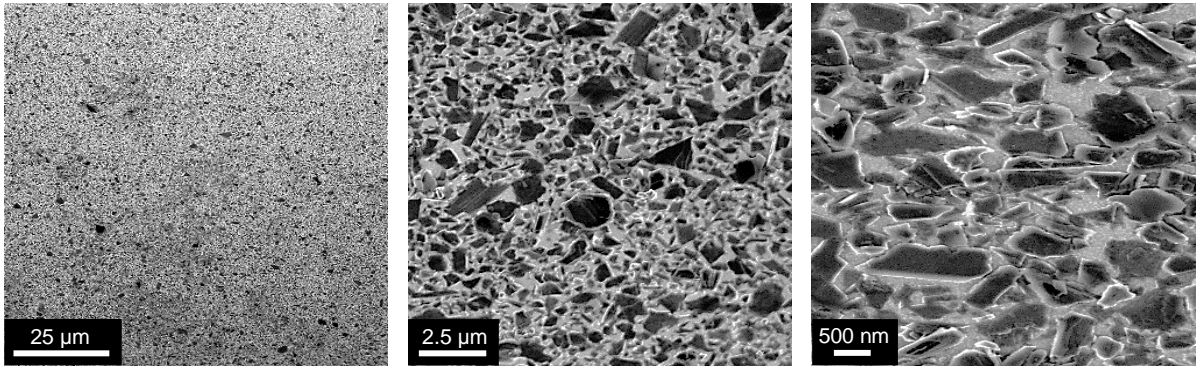

Small Nanosheets

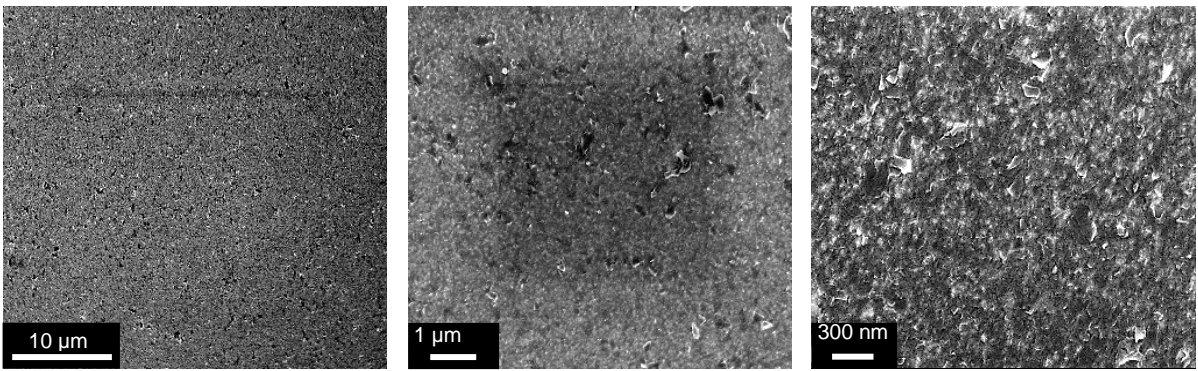

**Extended Data Fig 1** Scanning electron microscopy images of large (0.6-3k *g*) and small (10-30k *g*) nanosheets.

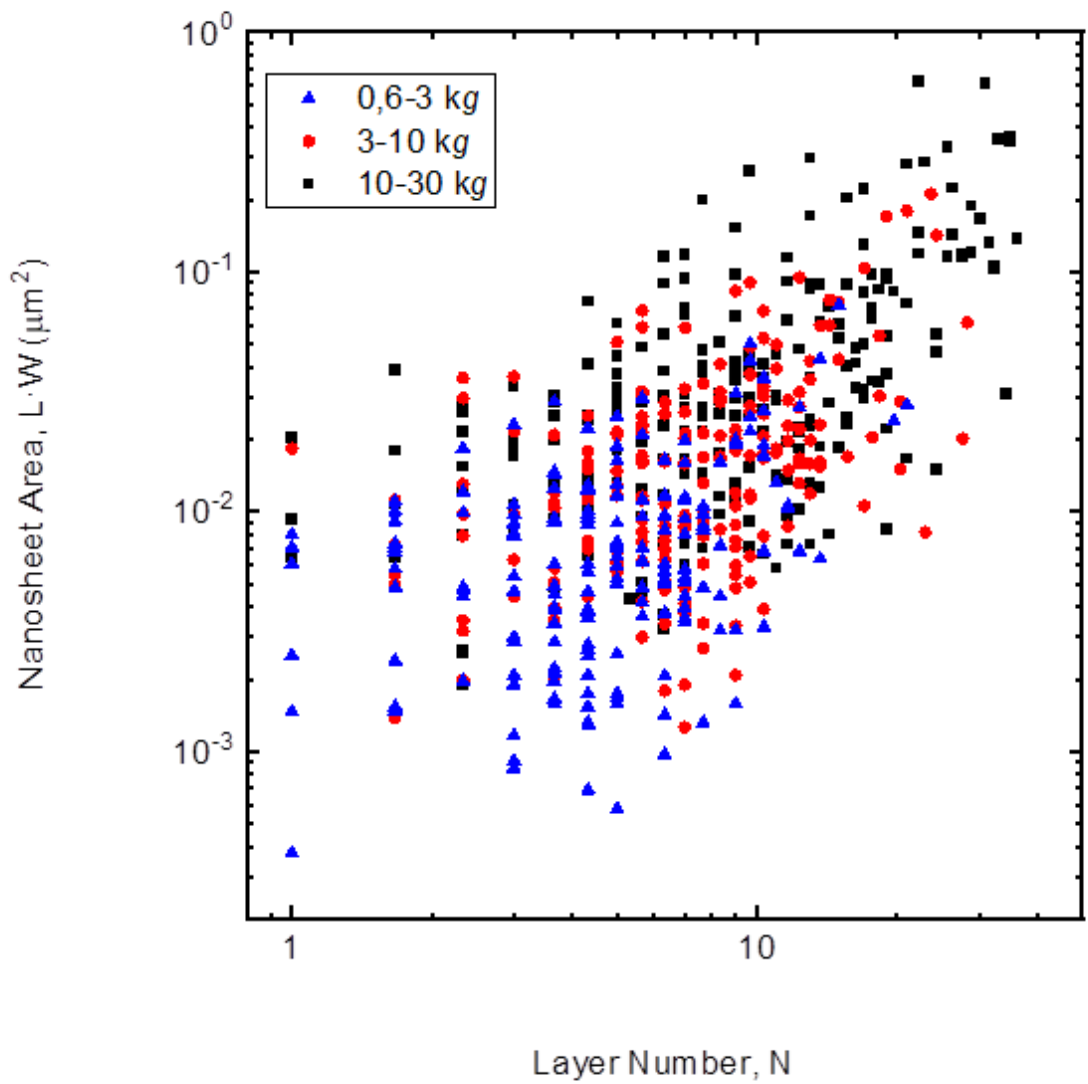

**Extended Data Fig 2 The ratio of lateral surface area to layer number for different LPE NiPS$_3$ flakes.** Blue, red and black symbols represent large, middle and small sized nanosheets.

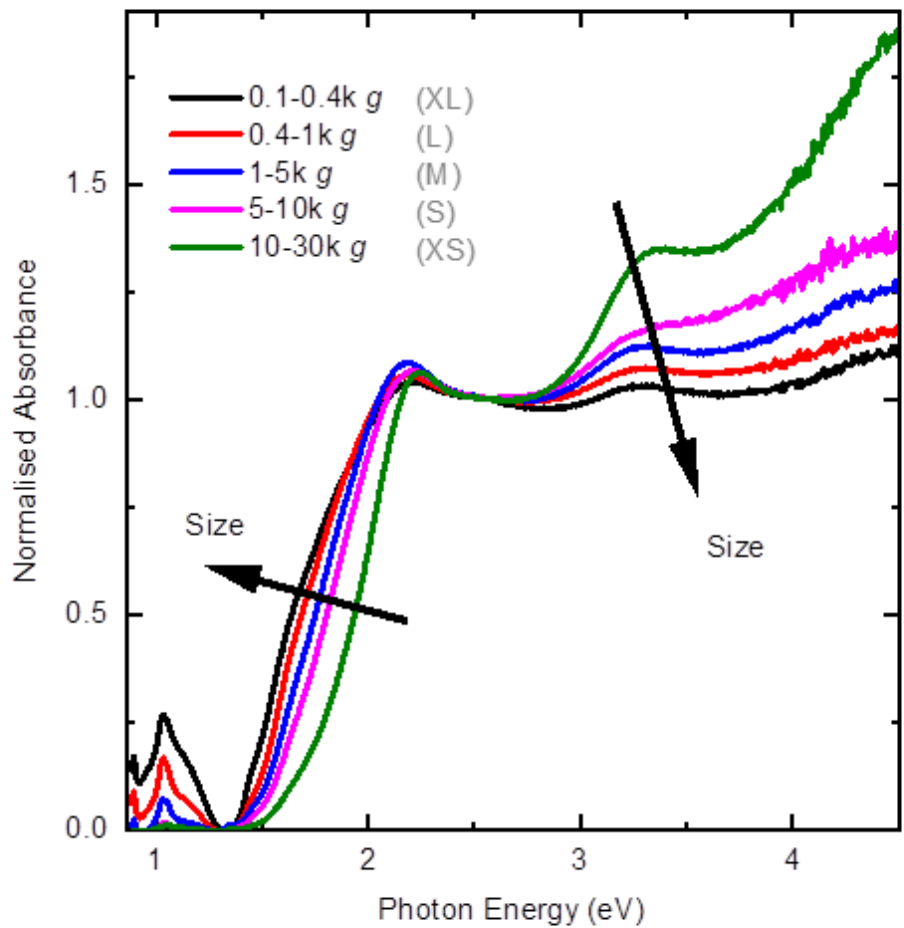

**Extended Data Fig 3 Room temperature absorbance of LPE NiPS$_3$ flakes of different lateral sizes.** Different curves represent absorbance spectra for samples prepared using different centrifugation parameters according to the legend of the graph. Arrows indicate the increase of the nanosheet size.

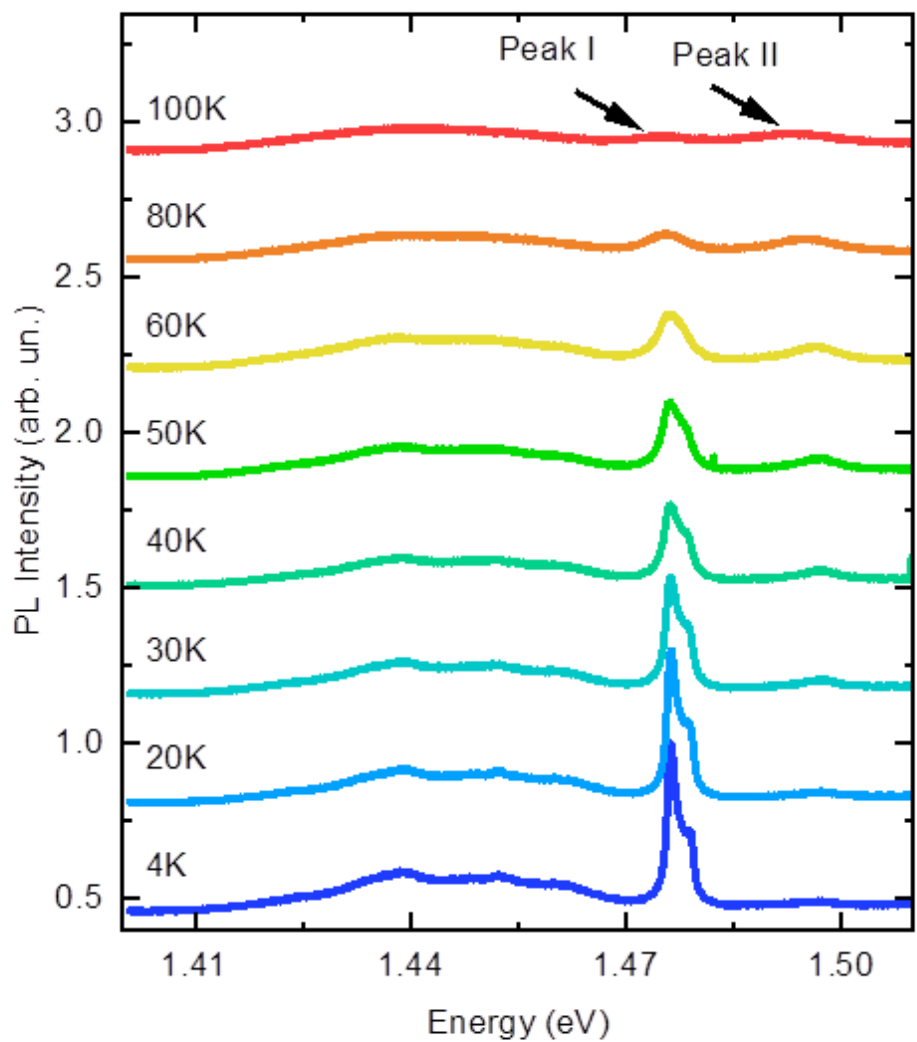

**Extended Data Fig 4** Extended temperature-dependent Photoluminescence spectra of large nanosheets (0.3-3k *g*) of LPE NiPS$_3$

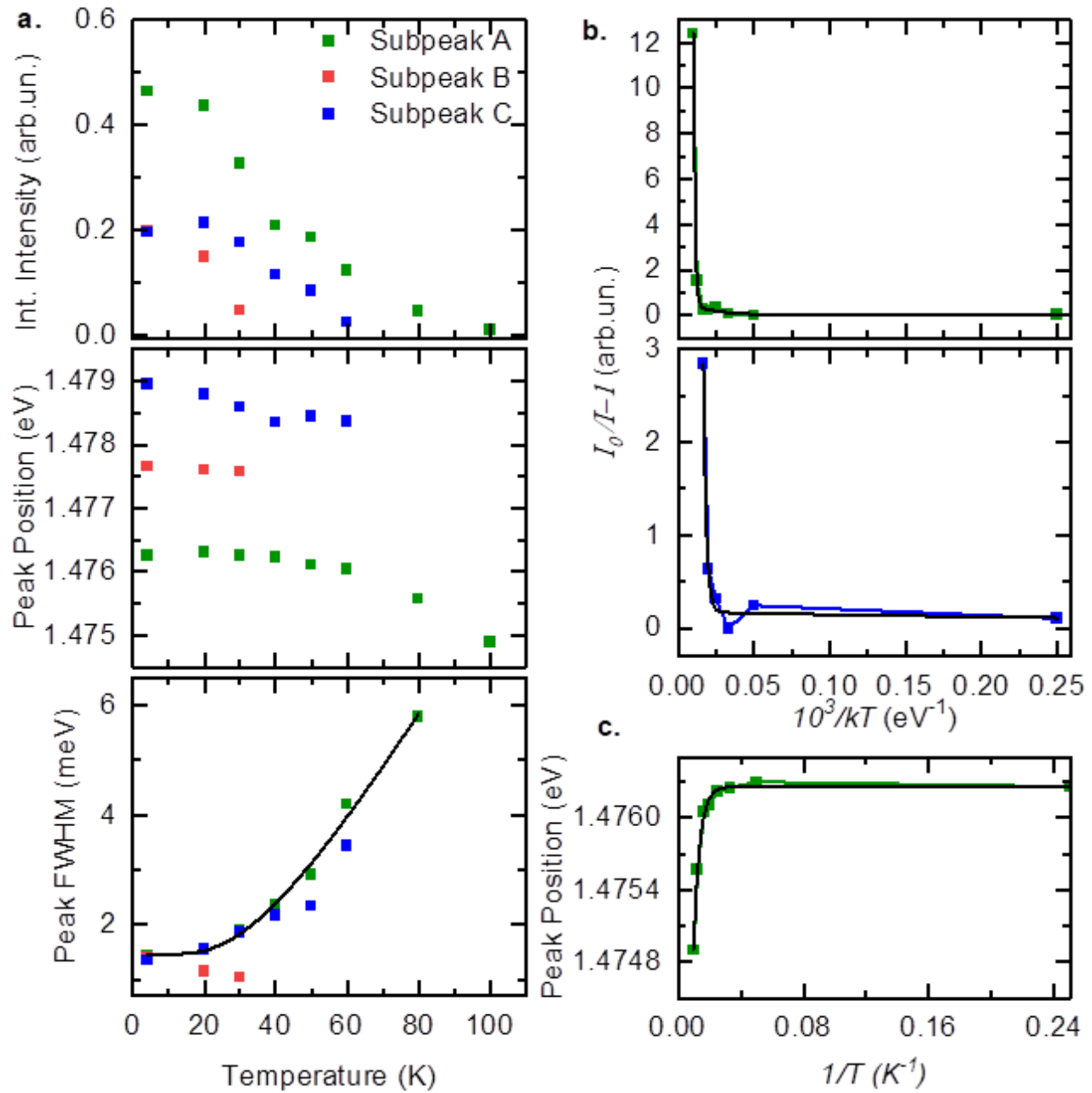

**Extended Data Fig 5 Temperature-dependent excitonic PL subpeak behavior of large (0.6-3k *g*) nanosheets of LPE NiPS$_3$. a,** Integrated subpeak intensity (upper graph), subpeak position (middle graph) and subpeak FWHM (lower graph) of excitonic PL. Green, red and blue dots represent subpeaks A, B and C as defined in the main text, respectively. The black line in the lower graph is the Bose Einstein fit (Formula 4) for the temperature dependence of subpeak A. **b,** Arrhenius plots (points) and fits according to Formula 2 (black lines) for the temperature dependent integrated intensity of Subpeak A (upper graph) and Subpeak C (lower graph). The upper graph was fitted with double exponential decay and the lower one with single exponential decay. **c,** Subpeak A position dependence on the inverse temperature and associated Bose-Einstein fit (black line) according to Formula 3. The parameters of all fits are indicated in Table 1.

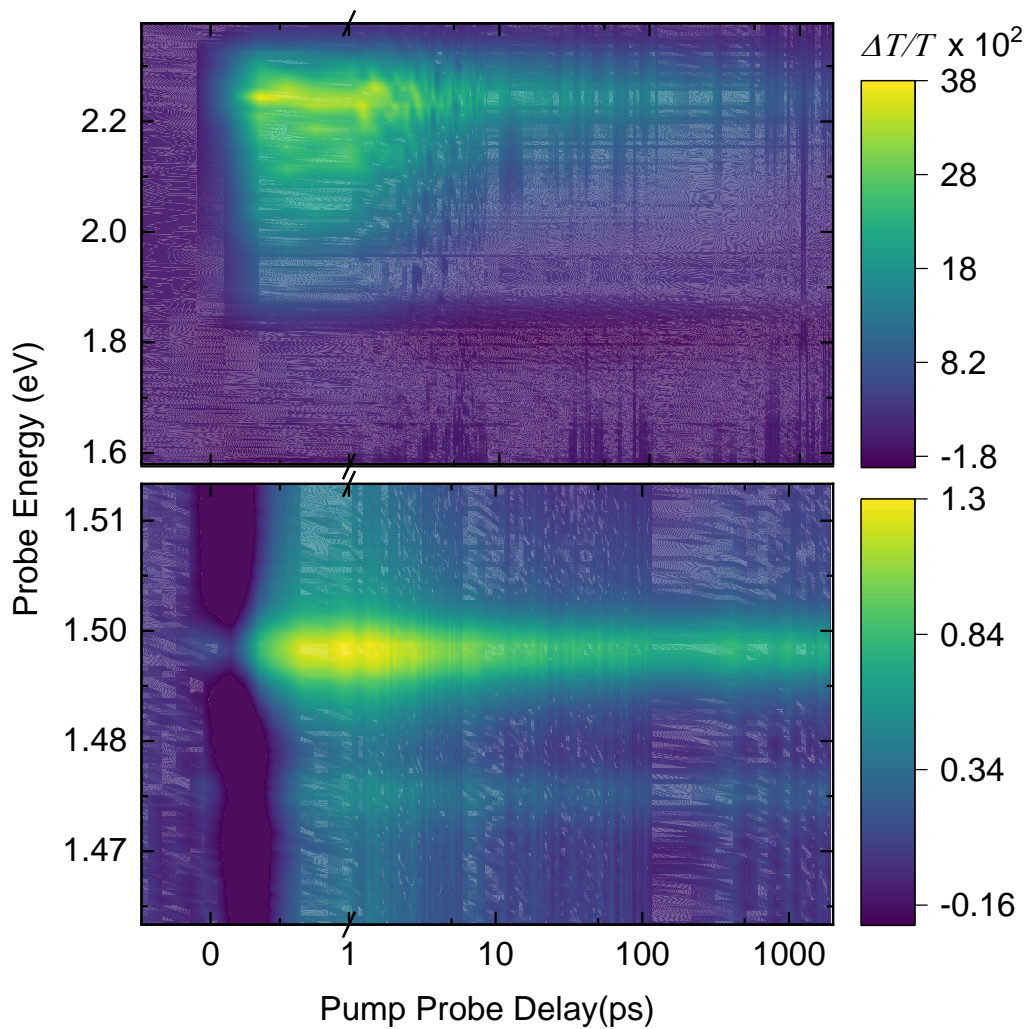

**Extended Data Fig 6 Transient absorption maps of large (0.6-3k *g*) nanosheets of LPE NiPS$_3$ measured at 4K.** ΔT/T map of LPE NiPS$_3$ absorption excited with 515 nm, 300 fs laser pulses, for pump-probe time delays and probe photon energies as shown.

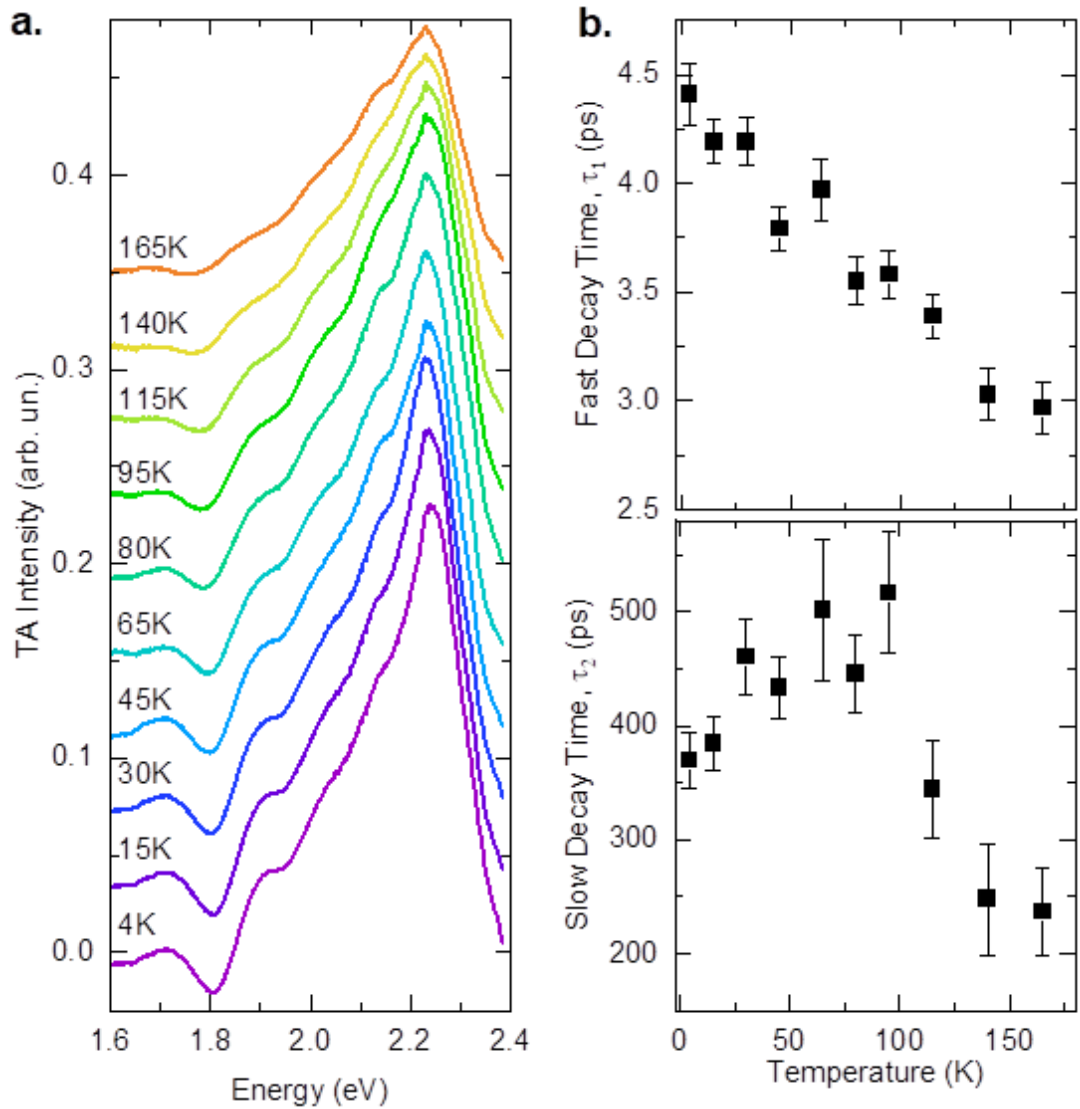

**Extended Data Fig 7 Transient absorption spectra of the ground state bleach (GSB) region of large (0.6-3k *g*) nanosheets of LPE NiPS$_3$. a,** Temperature-dependent dT/T spectra measured at temperatures from 4 K to 165 K in the region of GSB (1.6 eV to 2.4 eV) of LPE NiPS$_3$ averaged between 2 and 10 ps of pump-probe time delay. **b,** Temperature-dependent fast $\tau_1$ (upper graph) and slow $\tau_2$ (lower graph) decay time of GSB of LPE NiPS$_3$ extracted from the double exponential decay fit of the 2.18 eV - 2.3 eV probe data region in **a**. Error bars represent the error of the fit.

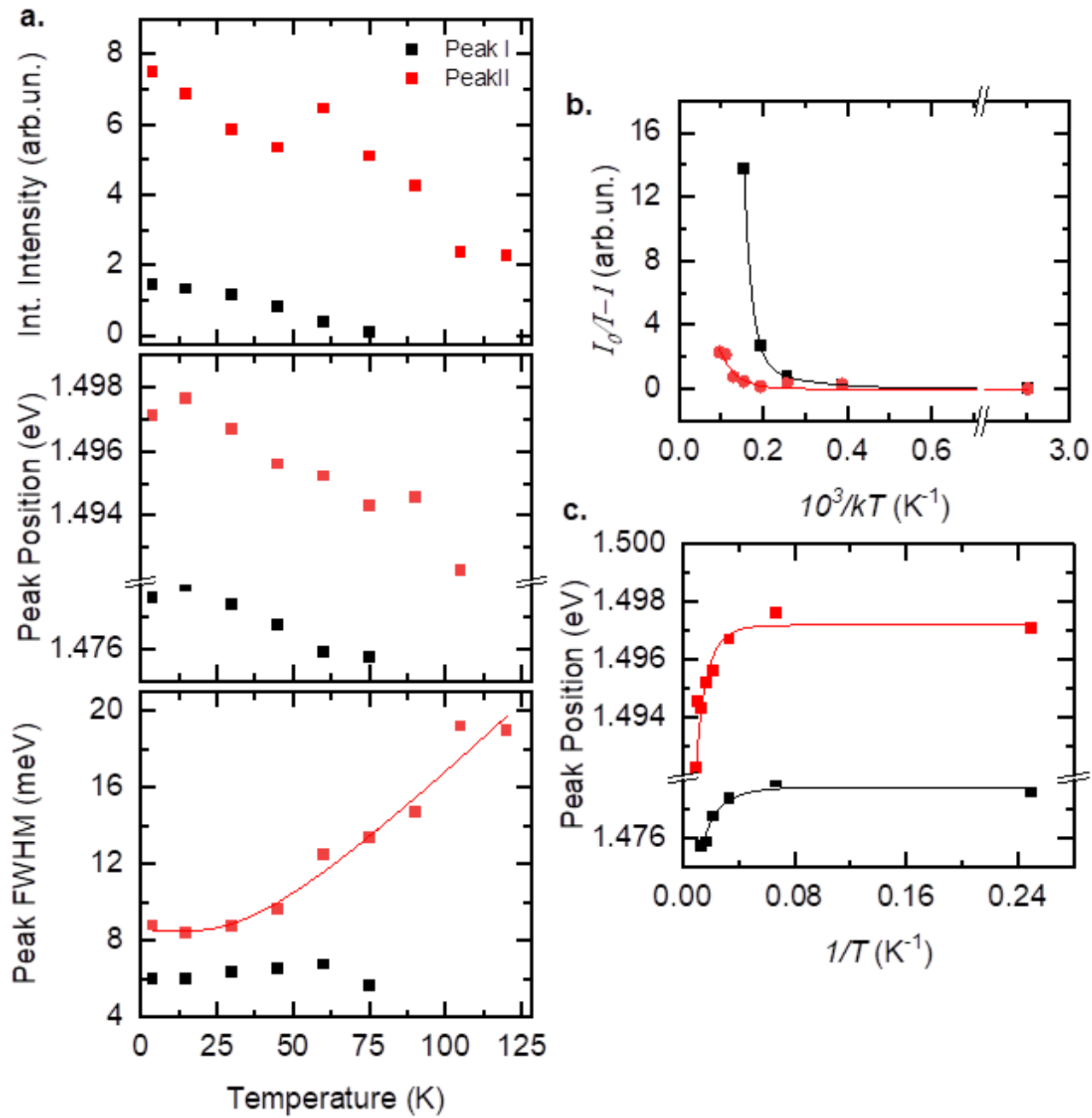

**Extended Data Fig 8 Temperature-dependent excitonic TA peak behavior of large (0.6-3k *g*) nanosheets of LPE NiPS$_3$ a,** Integrated peak intensity (upper graph), peak position (middle graph) and peak FWHM (lower graph) of excitonic TA. Black and red dots represent peaks I and II as defined in the main text respectively. The solid red line in the lower graph is the Bose-Einstein fit (Formula 4) for the temperature dependence of peak II. **b,** Arrhenius plots (points) and fits according to Formula 2 (solid lines) for the temperature dependent integrated intensity of peak I (black) and peak II (red). The peak II dependence was fitted with a double exponential decay and the peak I one with a single exponential decay. **c,** Peak I

(black) and II (red) position dependence on the inverse temperature and associated Bose-Einstein fit (solid lines) according to Formula 3. The parameters of all fits are indicated in Table 1.